\documentclass{article}

\usepackage{amssymb,amsmath,epsf}
\usepackage{pictexwd,dcpic}
\usepackage{wasysym}
\begin{document}

\title{The Canonical Lagrangian Approach To Three-Space General Relativity}         
\author{Vasudev Shyam and Madhavan Venkatesh\\ \\Centre for Fundamental Research and Creative Education,\\
Bangalore, India}        

\date{\today}

\maketitle
\begin{abstract}
We study the action for the three-space formalism of General Relativity, better known as the BF\'O (Barbour--Foster--\'O Murchadha) action, which is a square-root BSW (Baierlein--Sharp--Wheeler) action. In particular, we explore the (pre)symplectic structure by pulling it back via a Legendre map to the tangent bundle of the configuration space of this action. With it we attain the canonical Lagrangian vector field which generates the gauge transformations (3-diffeomorphisms) and the true physical evolution of the system. This vector field encapsulates all the dynamics of the system. We also discuss briefly the observables and perennials for this theory.  We then present a symplectic reduction of the constrained phase space.
\end{abstract}
\section{Introduction}       
In their `timeless' derivation of canonical general relativity via a theory which presupposes neither the relativity principle nor spacetime, Barbour \emph{et al} (\cite{bom1}) use a reparametrization invariant action, namely the Baierlein--Sharp--Wheeler (BSW) action, where the Lagrangian is integrated over an unphysical evolution parameter. We shall take a brief detour into their theory.
\subsection{The BSW action and `Relativity without Relativity'}    
The configuration space of the system considered here is Superspace, which is basically the set of all Riemannian metrics modulo a proper subgroup of the full 3 dimensional diffeomorphism group of a three manifold which, in the conventional ADM setting, is a space like hyper-surface embedded in spacetime. The topology is fixed so that the three manifold $\Sigma \cong S^{3}$. The ADM action for the standard (3+1) split is
\begin{equation} 
S=\int \textrm{d}t \textrm{d}^{3}x N \sqrt{\gamma}(R-K_{ab}K^{ab}-\textrm{tr}K^{2}).
\end{equation}
Now we replace $K_{ab}$ with $\frac{1}{2N}k_{ab}$ where
\begin{equation}k_{ab}= \dot{\gamma}_{ab}-\mathcal{L}_{\xi ^{a}}\gamma_{ab}. \end{equation}
The $\xi$ is an arbitrary vector field with respect to which the Lie Derivative acting on the metric represents the infinitesimal action of the 3 diffeomorphism group on the configuration space (which turns out to be equal to the shift of ADM gravity) and the over dot denotes differentiation with respect to an unphysical evolution parameter $\lambda $, and so the action now looks like
\begin{equation}S=\int \textrm{d}\lambda \textrm{d}^{3}x \sqrt{\gamma}\left[ NR-\frac{1}{4N}(k_{ab}k^{ab}-\textrm{tr}k^{2})\right]. \end{equation}
Varying with respect to $N$, we get
\begin{equation} N=\sqrt{\frac{k_{ab}k^{ab}-\textrm{tr}k^{2}}{4R}}. \end{equation}
Putting this back into the action, we find that
\begin{equation} S= \int \textrm{d}\lambda \textrm{d}^{3}x \sqrt{\gamma}\sqrt{R} \sqrt{T}, \end{equation}
where the `Kinetic Energy' term $T$ is
\begin{equation} T= G^{abcd}(\dot{\gamma}_{ab}-\mathcal{L}_{\xi^{a}}\gamma_{ab})(\dot{\gamma}_{cb}-\mathcal{L}_{\xi^{a}}\gamma_{cd}), \end{equation}
and the $G^{abcd}$ is the (inverse) DeWitt Supermetric.
\subsection{Some Preliminaries}
Here we shall discuss some of the mathematical preliminaries of our formalism. In general, a Hamiltonian system adheres to the following diagram:
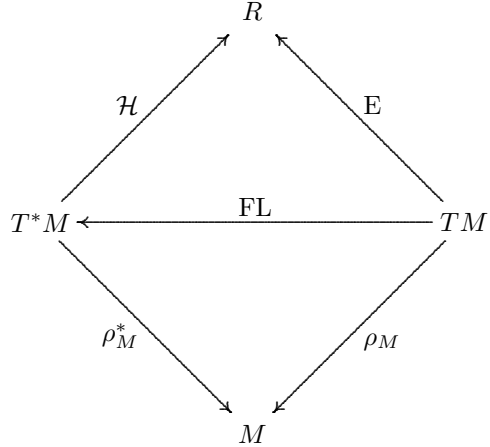
\[\begindc{0}[80]
\obj(1,2){$R$}
\obj(0,1){$T^{*}M$}
\obj(2,1){$TM$}
\obj(1,0){$M$}
\mor{$TM$}{$T^{*}M$}{FL}[-1,0]
\mor{$T^{*}M$}{$M$}{$\rho^{*}_{M}$}[0,0]
\mor{$TM$}{$M$}{$\rho_{M}$}
\mor{$T^{*}M$}{$R$}{$\mathcal{H}$}
\mor{$TM$}{$R$}{E}[-1,0]
\enddc\]
where $M$ is the configuration space. $FL$ is the fiber derivative between the tangent and the cotangent bundle. When it acts pointwise, it is identified with the standard Legendre transformation. The $\rho_{M}$s are the bundle projections from the tangent and the cotangent bundle down to $M$. $\mathcal{H}$ is the Hamiltonian and E is the energy function of the Lagrangian.
In the system we consider, the Lagrangian is the Barbour-Foster-\'O Murchadha Lagrangian
$L_{BFO}:TMet(\Sigma)\rightarrow R$,
given by
$$L_{BFO}=\int \textrm{d}^{3}x\sqrt{\gamma}\sqrt{T}\sqrt{R}.$$
For a presymplectic manifold $M$, which possesses a presymplectic form: $\Omega$, a vector field $$X_{f}\in TM$$ is said to be Hamiltonian if both
$$\mathcal{L}_{X_{f}}\Omega=0$$
$$\iota_{X_{f}}\Omega=\textrm{d}f$$
are satisfied, whereas it is locally Hamiltonian if only
$$\mathcal{L}_{X_{f}}\Omega=0$$
is satisfied globally. It should be noted that this terminology is carried over even to the tangent bundle of $Met(\Sigma)$ in this paper.

\section{The Lagrangian Presymplectic Potential}    
Now we shall derive the Lagrangian presymplectic potential, from which the presymplectic two form will follow. In this section we shall use the functional exterior derivative denoted by $\textrm{d}_{\delta}$ whose action is defined by
\begin{equation} \textrm{d}_{\delta}:\Omega^{p}(Met(\Sigma))\rightarrow \Omega^{p+1}(Met(\Sigma)).\end{equation}
Its properties are
$$\textrm{d}_{\delta}f=\frac{\delta f}{\delta x^{a}}\delta x^{a}$$ for $f \in \Omega^{0}(Met(\Sigma))$ and
$$\textrm{d}_{\delta}(\alpha \wedge \mu)=\textrm{d}_{\delta}\alpha \wedge \mu+ (-1)^{pq}\textrm{d}_{\delta}\mu \wedge \alpha$$
$$\textrm{d}_{\delta}\textrm{d}_{\delta}=0$$
$\forall \alpha \in \Omega^{p}(Met(\Sigma))$ and $\forall \mu \in \Omega^{q}(Met(\Sigma)).$
By Lagrangian presymplectic potential we mean the pullback via $FL$ of the Hamiltonian presymplectic potential i.e.
$$\Theta=(FL)^{*}\theta_{\mathcal{H}},$$
and, correspondingly, the presymplectic Lagrangian two form is given by
$$\Omega=(FL)^{*}\textrm{d}\theta_{\mathcal{H}}.$$
Therefore the Legendre map endows the tangent bundle of the configuration space too with presymplectic structure.
\subsection{The Constraint Submanifold of $TMet(\Sigma)$}
In the BF\'O approach, the Hamiltonian constraint of general relativity arises from the square root identity of the local square root action (thus the BF\'O authors recover infinitely many Hamiltonian constraints, i.e, one for each space point),
\begin{equation}  \frac{1}{\sqrt{\gamma}}(\pi^{ab}\pi_{ab}-\frac{1}{2}\textrm{tr}\pi^{2})-\sqrt{\gamma}R=0, \end{equation}
and the diffeomorphism constraint arises from their best matching method where
\begin{equation}\delta_{\xi}S_{BSW}=0, \end{equation}
which gives us
\begin{equation} -2\nabla_{a}\pi^{ab}=0. \end{equation}
The  surfaces  where these constraints are satisfied thus form the constraint submanifolds of $T^{*}Met(\Sigma)$. We shall attempt to get a similar constraint submanifold on $TMet(\Sigma)$. From the presymplectic algorithm on $T^{*}Met(\Sigma)$, we know that there exists an inclusion mapping from the final constraint submanifold $\Gamma_{\mathcal{H}}$ to $T^{*}Met(\Sigma)$, that is
\begin{equation}\Gamma_{\mathcal{H}}\stackrel{\pi_{\mathcal{H}}}\longrightarrow T^{*}Met(\Sigma).\end{equation}
And, by the pullback of the presymplectic form to $TMet(\Sigma)$,it too should possess a final constraint submanifold which shall be obtained from the following diagram: 
\[\begindc{0}[70]
\obj(0,1){$T^{*}Met(\Sigma)$}
\obj(2,1){$TMet(\Sigma)$}
\obj(0,0){$\Gamma_{H}$}
\obj(2,0){$\Gamma_{E}$}
\mor{$TMet(\Sigma)$}{$T^{*}Met(\Sigma)$}{$FL$}[-1,0]
\mor{$\Gamma_{H}$}{$T^{*}Met(\Sigma)$}{$\pi_{\mathcal{H}}$}
\mor{$\Gamma_{E}$}{$TMet(\Sigma)$}{$\pi_{E}$}[-1,0]
\mor{$\Gamma_{H}$}{$\Gamma_{E}$}{$\pi_{E}^{-1}\circ (FL)^{-1}\circ \pi_{\mathcal{H}}$}[-1,0]
\enddc\]
We know that
\begin{equation}(FL)^{-1}=F\mathcal{H},\end{equation}
since
\begin{equation}E=\mathcal{H}\circ FL=\pi_{ab}\dot{\gamma}^{ab}-L.\end{equation}
Therefore, the above mapping from the cotangent bundle constraint submanifold to that of the tangent bundle can be given by
\[\begindc{0}[50]
\obj(0,1){$\Gamma_{H}$}
\obj(2,1){$\Gamma_{E}$}
\mor{$\Gamma_{H}$}{$\Gamma_{E}$}{$\pi_{E}^{-1}\circ F\mathcal{H}\circ \pi_{\mathcal{H}}$}[1,0]
\enddc\]
And so, the restriction of the Lagrangian presymplectic potential to $TMet(\Sigma)$'s constraint submanifold is
\begin{equation}(\pi_{E}^{-1}\circ F\mathcal{H}\circ \pi_{\mathcal{H}})^{*}\theta_{\mathcal{H}}|_{\Gamma_{\mathcal{H}}}=\Theta|_{\Gamma_{E}}.\end{equation}
\subsection{The Derivation of the Potential}
In this section we present the derivation of the Lagrangian presymplectic potential from the BF\'O action. We have \begin{eqnarray}\delta S=0=\delta\int \textrm{d}\lambda \textrm{d}^{3}x \sqrt{\gamma}\sqrt{R} \sqrt{T}\end{eqnarray}
$$=>\int \textrm{d}\lambda \textrm{d}^{3}x \delta \sqrt{\gamma}\sqrt{R} \sqrt{T}+\int \textrm{d}\lambda \textrm{d}^{3}x \sqrt{\gamma}\delta\sqrt{R} \sqrt{T}+\int \textrm{d}\lambda \textrm{d}^{3}x \sqrt{\gamma}\sqrt{R} \delta\sqrt{T}.$$
Expanding the variation of the third term explicitly, and integrating by parts, we attain
\begin{eqnarray}\int \textrm{d}^{3}x \sqrt{\frac{\gamma R}{4T}}\frac{\delta T}{\delta \dot{\gamma}_{ab}}\delta \gamma^{ab}|_{\lambda}-\int \textrm{d}^{3}x \textrm{d}\lambda[ \delta \gamma_{ab}\left[ \sqrt{\frac{\gamma R}{4T}}\frac{\textrm{d}}{\textrm{d} \lambda}\left(\frac{\delta T}{\delta \dot{\gamma}_{ab}}\right)\right]+\delta \sqrt{\gamma}\sqrt{R} \sqrt{T}+ \end{eqnarray}
$$\sqrt{\gamma}\delta\sqrt{R} \sqrt{T}].$$ 
The second term shall vanish by the Euler-Lagrange equations, but our interest is with the first term
\begin{equation}\int \textrm{d}^{3}x \sqrt{\frac{\gamma R}{4T}}\frac{\delta T}{\delta \dot{\gamma}_{ab}}\delta \gamma^{ab}|_{\lambda}=\int \textrm{d}^{3}x \sqrt{\frac{\gamma R}{4T}}G^{abcd}[\dot{\gamma}_{ab}-\mathcal{L}_{\xi^{a}}\gamma_{ab}]\delta \gamma_{ab}=\Theta|_{\Gamma_{E}}. \end{equation}
where $\Theta|_{\Gamma_{E}}$ is the presymplectic potential constrained to the constraint submanifold $\Gamma_{E}$. The presymplectic two form is thus
\begin{equation}\textrm{d}_{\delta}\Theta|_{\Gamma_{E}}=\Omega|_{\Gamma_{E}}=\int \textrm{d}^{3}x (\textrm{d}_{\delta}\sqrt{\frac{\gamma R}{4T}}k_{ab}\wedge \textrm{d}_{\delta}\gamma^{ab}).\end{equation}
\section{Dynamics and the Presymplectic Equation}
Since the presymplectic two form is degenerate, we can only derive a Canonical locally Hamiltonian vector field called the Lagrangian vector field which satisfies
\begin{eqnarray}\mathcal{L}_{X}\textrm{d}_{\delta}\Theta|_{\Gamma_{E}}=0\\
=> \iota_{X}\textrm{d}^{2}_{\delta}\Theta|_{\Gamma_{E}}+\textrm{d}_{\delta}\iota_{X}\textrm{d}_{\delta}\Theta|_{\Gamma_{E}}=0.
\end{eqnarray}
The first term vanished by $\textrm{d}^{2}_{\delta}=0$, 
\begin{eqnarray}\textrm{d}_{\delta}\iota_{X}\textrm{d}_{\delta}\Theta|_{\Gamma_{E}}=0\\
\iota_{X}\textrm{d}_{\delta}\Theta|_{\Gamma_{E}}=\textrm{d}_{\delta}E|_{\Gamma_{E}}. \end{eqnarray}
This is due to the Poincare Lemma, and the E is the Energy Functional, which is given by
$$E|_{\Gamma_{E}}=\sqrt{\frac{\gamma R}{4T}}G^{abcd}k_{ab}k_{cd}-\sqrt{\frac{T}{4\gamma R}}R,$$
and so
$$X|_{\Gamma_{E}}=\pi_{E*}X=X_{E},$$
where $\pi_{E}$ is the mapping from the constraint surface to the total phase space.
From this, we attain the expression for the (locally Hamiltonian) Lagrangian  vector field 
\begin{eqnarray}
X_{E}=\sqrt{\frac{4T}{\gamma R}}[k_{ab}+\mathcal{L}_{\xi^{a}}\gamma_{ab}]\frac{\delta}{\delta \gamma_{ab}}- [\sqrt{\frac{\gamma T}{4R}}(R^{ab}-\frac{1}{2}\gamma^{ab}R)-\sqrt{\frac{T}{\gamma R}}(k^{ac}k_{c}^{b}-\frac{1}{2}k k^{ab})\\
-\nabla^{a}\sqrt{\frac{T}{4\gamma R}}\nabla^{b}\sqrt{\frac{T}{4\gamma R}}+\gamma^{ab}\nabla^{2}\sqrt{\frac{T}{4\gamma R}}+\mathcal{L}_{\xi^{a}}\sqrt{\frac{T}{4\gamma R}}k^{ab}]\frac{\delta}{\delta k^{ab}}.\end{eqnarray}
The canonical Lagrangian vector field belongs to $TTMet(\Sigma)$. On $TT^{*}Met(\Sigma)$ the Hamiltonian vector field would satisfy the presymplectic equation
$$(X_{\mathcal{H}})^{\flat}|_{\Gamma_{\mathcal{H}}}=0,$$
where
$$\flat:TT^{*}Met(\Sigma)\rightarrow T^{*}T^{*}Met(\Sigma),$$
acts as a local isomorphism when restricted to the constraint submanifold. Now, in order to attain the same for our present formalism, we refer to the diagram:
\[\begindc{0}[70]
\obj(0,2){$T^{*}T^{*}Met(\Sigma)$}
\obj(2,2){$T^{*}TMet(\Sigma)$}
\obj(0,1){$TT^{*}Met(\Sigma)$}
\obj(2,1){$TTMet(\Sigma)$}
\obj(0,0){$T^{*}Met(\Sigma)$}
\obj(2,0){$TMet(\Sigma)$}
\mor{$TT^{*}Met(\Sigma)$}{$TTMet(\Sigma)$}{$TFL$}[1,0]
\mor{$T^{*}Met(\Sigma)$}{$TMet(\Sigma)$}{$FL$}[1,0]
\mor{$T^{*}Met(\Sigma)$}{$TT^{*}Met(\Sigma)$}{$X_{\mathcal{H}}$}[1,0]
\mor{$TMet(\Sigma)$}{$TTMet(\Sigma)$}{$X_{E}$}[-1,0]
\mor{$T^{*}T^{*}Met(\Sigma)$}{$T^{*}TMet(\Sigma)$}{$T^{*}FL$}[1,0]
\mor{$TT^{*}Met(\Sigma)$}{$T^{*}T^{*}Met(\Sigma)$}{$\flat$}[1,0]
\mor{$TTMet(\Sigma)$}{$T^{*}TMet(\Sigma)$}{$T^{*}FL\circ \flat \circ(TFL)^{-1}$}[-1,0]
\enddc\]
And so we obtain the presymplectic equation for the Canonical Lagrangian vector field:
\begin{equation}(X_{E})^{(T^{*}FL\circ \flat \circ(TFL)^{-1})}|_{\Gamma_{E}}=0,\end{equation}
which for the sake of brevity shall be written as
\begin{equation}(X_{E})^{\quarternote}|_{\Gamma_{E}}=0,\end{equation}
where
$$\quarternote:=(T^{*}FL\circ \flat \circ(TFL)^{-1}),$$
whose action on vector fields of $TTMet(\Sigma)$ is defined as
$$Z^{\quarternote}=\Omega(Z).$$
So, in totality, the geometry of the dynamical system described by the BF\'O action is given by:
\[\begindc{0}[80]
\obj(0,4){$T^{*}T^{*}Met(\Sigma)$}
\obj(2,4){$T^{*}TMet(\Sigma)$}
\obj(0,3){$TT^{*}Met(\Sigma)$}
\obj(2,3){$TTMet(\Sigma)$}
\obj(0,2){$T^{*}Met(\Sigma)$}
\obj(2,2){$TMet(\Sigma)$}
\obj(1,1){$Met(\Sigma)$}
\obj(0,0){$\Gamma_{H}$}
\obj(2,0){$\Gamma_{E}$}
\mor{$TT^{*}Met(\Sigma)$}{$TTMet(\Sigma)$}{$TFL$}[1,0]
\mor{$T^{*}Met(\Sigma)$}{$TMet(\Sigma)$}{$FL$}[1,0]
\mor{$T^{*}Met(\Sigma)$}{$TT^{*}Met(\Sigma)$}{$X_{\mathcal{H}}$}[1,0]
\mor{$TMet(\Sigma)$}{$TTMet(\Sigma)$}{$X_{E}$}[-1,0]
\mor{$T^{*}T^{*}Met(\Sigma)$}{$T^{*}TMet(\Sigma)$}{$T^{*}FL$}[1,0]
\mor{$TT^{*}Met(\Sigma)$}{$T^{*}T^{*}Met(\Sigma)$}{$\flat$}[1,0]
\mor{$TTMet(\Sigma)$}{$T^{*}TMet(\Sigma)$}{$\quarternote$}[-1,0]
\mor{$T^{*}Met(\Sigma)$}{$Met(\Sigma)$}{$\rho^{*}_{Met(\Sigma)}$}[0,0]
\mor{$TMet(\Sigma)$}{$Met(\Sigma)$}{$\rho_{Met(\Sigma)}$}
\mor{$\Gamma_{H}$}{$T^{*}Met(\Sigma)$}{$\pi_{\mathcal{H}}$}
\mor{$\Gamma_{E}$}{$TMet(\Sigma)$}{$\pi_{E}$}[-1,0]
\mor{$\Gamma_{H}$}{$\Gamma_{E}$}{$\pi_{E}^{-1}\circ (FL)^{-1}\circ \pi_{\mathcal{H}}$}[1,0]
\enddc\]
\subsection{Flows Of The Canonical Vector Field}
The canonical flow of a function on $TMet{\Sigma}$ is given by the solution to the following Cauchy problem
\begin{equation}f^{0}_{E}[z(\gamma_{ab},\sqrt{\frac{\gamma R}{4T}}k_{ab})]=z(\gamma_{ab},\sqrt{\frac{\gamma R}{4T}}k_{ab})\end{equation}
\begin{equation}\frac{\textrm{d}}{\textrm{d}\lambda}f^{\lambda}_{E}[z(\gamma_{ab},\sqrt{\frac{\gamma R}{4T}}k_{ab})]|_{\lambda=0}=X_{E}[z(\gamma_{ab},\sqrt{\frac{\gamma R}{4T}}k_{ab})].\end{equation}
This can be solved by
\begin{equation}f^{\lambda}_{E}[z(\gamma_{ab},\sqrt{\frac{\gamma R}{4T}}k_{ab})]=\sum_{n=0}^{\infty}\frac{\lambda^{n}}{n!}X^{n}_{E}[z(\gamma_{ab},\sqrt{\frac{\gamma R}{4T}}k_{ab})].\end{equation}
So,
\begin{equation}f^{\lambda}_{E}[z(\gamma_{ab},\sqrt{\frac{\gamma R}{4T}}k_{ab})]=z(\gamma_{ab},\sqrt{\frac{\gamma R}{4T}}k_{ab};\lambda).\end{equation}
Now the entire evolution dynamics of the RWR theory is given by the Euler Lagrange equation for $\gamma_{ab}$:
\begin{eqnarray}\frac{\textrm{d}p^{ab}}{\textrm{d}\lambda}=\sqrt{\frac{\gamma T}{4R}}(R^{ab}-\frac{1}{2}\gamma^{ab}R)-\sqrt{\frac{T}{\gamma R}}(p^{ac}p_{c}^{b}-\frac{1}{2}pp^{ab})\\
-\nabla^{a}\sqrt{\frac{T}{4\gamma R}}\nabla^{b}\sqrt{\frac{T}{4\gamma R}}+\gamma^{ab}\nabla^{2}\sqrt{\frac{T}{4\gamma R}}+\mathcal{L}_{\xi^{a}}p^{ab},\end{eqnarray}
which, in our formalism, is nothing but
\begin{eqnarray}\frac{\textrm{d}}{\textrm{d}\lambda}\left[\sqrt{\frac{\gamma R}{4T}}k_{ab}(\lambda)\right]|_{\lambda=0}=\sqrt{\frac{\gamma T}{4R}}(R^{ab}-\frac{1}{2}\gamma^{ab}R)-\sqrt{\frac{T}{\gamma R}}(k^{ac}k_{c}^{b}-\frac{1}{2}k k^{ab})-\\ \nabla^{a}\sqrt{\frac{T}{4\gamma R}}\nabla^{b}\sqrt{\frac{T}{4\gamma R}}+\gamma^{ab}\nabla^{2}\sqrt{\frac{T}{4\gamma R}}+\mathcal{L}_{\xi^{a}}\sqrt{\frac{T}{4\gamma R}}k^{ab},\end{eqnarray}
which is but the flow equation for the velocity conjugate to the metric on $TMet(\Sigma)$.
\subsection{The Splitting of the Canonical Vector field}
The algebra of canonical Lagrangian vector fields derived in the previous section is given by
\begin{equation}\left\{X:\mathcal{L}_{X}\textrm{d}_{\delta}\Theta|_{\Gamma_{E}}=0;X \in TTMet(\Sigma)\right\}=\natural(\Omega^{1}_{\scriptsize\textrm{closed}}(Met(\Sigma))).\end{equation}
(Here $\Omega^{1}_{\scriptsize\textrm{closed}}(Met(\Sigma)$ is the set of all closed one forms on the space of three metrics),and the $\natural$ refers to the inverse of the $\quarternote$ map. As this map is an isomorphism (when restricted to $\Gamma_{E}$), so we find that the canonical Lagrangian vector field admits the split
\begin{equation}(\textrm{d}_{\delta}\Theta)^{\natural}|_{\Gamma_{E}}=\mathcal{E}v_{\sqrt{\frac{T}{4R}}}+\mathcal{G}_{\xi^{a}},\end{equation}
where $\mathcal{E}v_{\sqrt{\frac{T}{4R}}}$ is the vector field responsible for true physical evolution of the 
system and $\mathcal{G}_{\xi^{a}}$ is the vector field which represents the infinitesimal action of the three-Diffeomorphism group (rather its proper subgroup $Diff_{F}(\Sigma)$). It is given by
\begin{equation}\mathcal{G}_{\xi^{a}}= -[\mathcal{L}_{\xi^{a}}\sqrt{\frac{T}{4\gamma R}}k^{ab}]\frac{\delta}{\delta k^{ab}}+[\mathcal{L}_{\xi^{a}}\gamma_{ab}]\frac{\delta}{\delta \gamma_{ab}},\end{equation}
and the evolutionary vector field is
\begin{eqnarray}\mathcal{E}v_{\sqrt{\frac{T}{4R}}}= [\sqrt{\frac{4T}{\gamma R}}k_{ab}]\frac{\delta}{\delta \gamma_{ab}}- [\sqrt{\frac{\gamma T}{4R}}(R^{ab}-\frac{1}{2}\gamma^{ab}R)-\sqrt{\frac{T}{\gamma R}}(k^{ac}k_{c}^{b}-\frac{1}{2}k k^{ab})\\
-\nabla^{a}\sqrt{\frac{T}{4\gamma R}}\nabla^{b}\sqrt{\frac{T}{4\gamma R}}+\gamma^{ab}\nabla^{2}\sqrt{\frac{T}{4\gamma R}}]\frac{\delta}{\delta k^{ab}}.\end{eqnarray}
\subsection{Observables and Perennials}
There exists a natural identification of the observables and perennials of this system given by the following  distributions on phase space:
We define the set of perennials as
\begin{equation}  \mathcal{P}=\mathcal{O}\cap \mathcal{D}=\left\{f(\gamma_{ab},\sqrt{\frac{\gamma R}{4T}}k_{ab};\lambda) :X_{E}\left[f(\gamma_{ab},\sqrt{\frac{\gamma R}{4T}}k_{ab};\lambda)\right]=0\right\}.\end{equation}
The set of observables is defined by the set
\begin{equation}\mathcal{O}=\left\{g(\gamma_{ab},\sqrt{\frac{\gamma R}{4T}}k_{ab};\lambda):\mathcal{G}_{\xi^{a}}\left[g(\gamma_{ab},\sqrt{\frac{\gamma R}{4T}}k_{ab};\lambda)\right]=0\right\}.\end{equation}
And  $\mathcal{D}$ is
\begin{equation}\mathcal{D}=\left\{h(\gamma_{ab},\sqrt{\frac{\gamma R}{4T}}k_{ab};\lambda):\mathcal{E}v_{\sqrt{\frac{T}{4R}}}\left[h(\gamma_{ab},\sqrt{\frac{\gamma R}{4T}}k_{ab};\lambda)\right]=0\right\}.\end{equation}
It is known that for a vacuum gravitational field, any such observable must be a highly non-local quantity. The ADM mass in the case of asymptotically flat scenarios is an example.
\section{Symmetry and Reduction}
We shall now attain the reduced phase space of the theory on symplectic reduction of the velocity phase space.
\subsection{The Moment Map Associated with the Symmetry}
The Lie algebra of the group acting on phase space is $\mathfrak{g}=\mathfrak{diff}_{F}(\Sigma).$ We need to find a moment map
\begin{equation}\mu:\Gamma_{E} \rightarrow \mathfrak{diff}^{*}_{F}(\Sigma).\end{equation}
(Here $\mathfrak{diff}^{*}_{F}(\Sigma)$ is the dual of $\mathfrak{diff}_{F}(\Sigma)$.)
This moment map is $Diff_{F}(\Sigma)$ equivariant. Thus we consider the level set preserved by $Diff_{F}(\Sigma)$
\begin{equation}\Gamma_{E}(0)=\left\{m \in\Gamma_{E}|\mu(m)=0\right\}.\end{equation}
These level sets foliate the constraint submanifold into gauge orbits. It isn't hard to see that the phase space function satisfying these conditions for this particular system is
\begin{equation} 2\nabla_{a}\left( \sqrt{\frac{\gamma R}{T}}G^{abcd}[\dot{\gamma}_{cd}-\mathcal{L}_{\xi^{c}}\gamma_{cd}]\right)=\phi(\xi).\end{equation}
This is nothing but the diffeomorphism constraint. Its $Diff_{F}(\Sigma)$ equivariance is shown via
\begin{equation}\iota_{\mathcal{G}_{\xi^{a}}}\phi(\zeta^{b})=\phi([\xi^{a},\zeta^{b}]).\end{equation}
This is analogous to the best matching procedure of the original BF\'O approach.
\subsection{The Reduction}
In this section the (pre-)symplectic reduction of the phase space by the symmetry is presented. We begin by applying the Marsden--Weinstein Reduction theorem (which we truncate accordingly for the presymplectic case). Let
$\tilde{\Gamma}_{E}:=\Gamma_{E}(0)/Diff_{F}(\Sigma)$. $\Gamma_{E}(0)$ could also be written as  $\mu^{-1}(0)$. Now, the reduction theorem tells us that there is an inclusion $$i:\mu^{-1}(0)\rightarrow \Gamma_{E}$$, and another map $$j:\mu^{-1}(0)\rightarrow \tilde{\Gamma}_{E},$$ for which there exists a presymplectic form $\omega \in \Omega^{2}(\tilde{\Gamma}_{E}),$ so that
$$i^{*}\Omega=j^{*}\omega.$$
Thus we see that after reduction we go to $$\tilde{\Gamma}_{E}\equiv \Gamma_{E}(0)//Diff_{F}(\Sigma),$$
and
\begin{equation}\Gamma_{E}(0)//Diff_{F}(\Sigma)\subset TS_{F}(\Sigma)=T(Met(\Sigma)/Diff_{F}(\Sigma)).\end{equation}
The symmetry group here is the proper subgroup of the Diffeomorphism group where the group action fixes a preferred point $\infty \in \Sigma$ and the tangent space at that point i.e
\begin{equation} Diff_{F}(\Sigma)=\left\{ \phi \in Diff(\Sigma)| \phi(\infty)=\infty,\phi_{*}(\infty)=Id|_{T_{\infty}\Sigma} \right\}. \end{equation}
This ensures that the action of this group is free and proper when $\Sigma$ is connected and compact, which is true for the topology of $S^{3}$ that we have fixed, and so the reduced phase space is ensured to be a manifold (See \cite{giu1} for further details). It may seem strange that even after reduction, but this is due to the fact that we have only reduced by the group of diffeomorphisms, for in the interpretation of this theory, it is the only constraint of the theory that generates gauge transformations, and the Hamiltonian constraint generates true dynamical evolution. But, with regard to the velocity phase space, the lack of a true Hamiltonian is still problematic as it prevents one from attaining a strongly non degenerate symplectic structure on the phase space, and so $\omega$ is still weakly non degenerate, and it satisfies the equation
$$\omega(\mathcal{E}v_{\sqrt{\frac{T}{4R}}})|_{\tilde{\Gamma}_{E}}=0.$$
Also, as the Hamiltonian constraint is not an equivariant moment map, symplectic reduction akin to that which has been carried out in this section will not be feasible for it.

\section{Concluding remarks}
In this paper, we have shown that the  dynamics of Three Space General Relativity can be dealt with on the tangent bundle of the space of Riemannian metrics by studying the presymplectic structure associated to it. Also, the first principles and the action of the BF\'O approach remain intact and the presymplectic two form is derived out of the BF\'O action. We find that the notion of observables and perennials arises naturally without considering the Poisson brackets of the functions with the constraints.We have also shown that the moment map used in the reduction procedure comes directly out of best matching. Even though this formalism relies on nothing but the first principles of the `Relativity Without Relativity' approach, a purely Hamiltonian framework is  still necessary and this shall be the subject of future papers.
\section{Acknowledgments}
This work was  carried out at the Center For Fundamental Research And Creative Education, Bangalore, India, under the guidance of Dr.B.S Ramachandra whom we wish acknowledge. We would also like to sincerely thank Julian Barbour for very valuable advice.

\end{document}